\documentclass[aps,prl,twocolumn,letter]{revtex4-1}
\usepackage{graphicx}
\usepackage{float}
\usepackage{amsmath}
\usepackage{color}

\newcommand{\be}{\begin{equation}}
\newcommand{\ee}{\end{equation}}

\begin{document}

\title{Direct measurement of the ballistic motion of a freely floating colloid in Newtonian and viscoelastic fluids }
\date{\today}
\author{Andrew P. Hammond, Eric I. Corwin}
\affiliation{Materials Science Institute and Department of Physics, University of Oregon, Eugene, Oregon 97403}

\begin{abstract}
A thermal colloid suspended in a liquid will transition from a short time ballistic motion to a long time diffusive motion.  However, the transition between ballistic and  diffusive motion is highly dependent on the properties and structure of the particular liquid.  We directly observe a free floating tracer particle's ballistic motion and its transition to the long time regime in both a Newtonian fluid and a viscoelastic Maxwell fluid.  We examine the motion of the free particle in a Newtonian fluid and demonstrate a high degree of agreement with the accepted Clercx-Schram model for motion in a dense fluid.  Measurements of the functional form of the ballistic-to-diffusive transition provide direct measurements of the temperature, viscosity, and tracer radius.  We likewise measure the motion in a viscoelastic Maxwell fluid and find a significant disagreement between the theoretical asymptotic behavior and our measured values of the microscopic properties of the fluid. We observe a greatly increased effective mass for a freely moving particle and a decreased plateau modulus.

\end{abstract}

\maketitle

At very short time and length scales the diffusive motion of a Brownian particle breaks down into a series of individual ballistic flights.  The functional form of this transition is controlled by the microscopic structure and behavior of the fluid.  The microscopic time and length scales for the ballistic motion are so small that direct measurements have only recently become possible \cite{lukic_direct_2005, huang_direct_2011, franosch_resonances_2011, kheifets_observation_2014}. These experiments have used optical traps to confine a test particle within a harmonic well, allowing a high precision measurement of the short time motion but at the cost of a loss of information about the crossover to longer time behavior that is indicative of the microscopic structure of the fluid. Furthermore, laser traps by their nature create a harmonic potential energy well for the motion of the particle and thus function similarly to an elastic term in a viscoelastic fluid.  As such, it can be difficult to deconvolve the effect of the trap from the effects of the elastic component of the fluid.  Indeed, all studies of viscoelastic fluids known to the authors don't address the ballistic regime. Here we avoid the limitations and contaminations caused by the use of a laser trap and present direct measurements of the full transition away from ballistic motion for a freely moving colloid suspended in simple Newtonian and viscoelastic Maxwell fluids.  These measurements are achieved in an interaction free manner using a high speed camera, intense illumination, and an accurate tracking algorithm \cite{parthasarathy_rapid_2012}.  These measurements allows us to unambiguously distinguish between microscopic models for dense fluid thermal motion \cite{lemons_paul_1997, uhlenbeck_theory_1930, zwanzig_hydrodynamic_1970,widom_velocity_1971, hinch_application_1975, clercx_brownian_1992,van_zanten_brownian_2000, wilhelm_rotational_2003, atakhorrami_short-time_2008,  raikher_brownian_2013} and provide a hithertofore impossible glimpse into the fundamental behavior of thermal fluids.  In a simple Newtonian fluid, our measurement is in close correspondence with analytic predictions.  By fitting our data we can directly measure the constants of motion as well as a first principles measurement of the temperature of the fluid.  Having proven the validity of this method, we experimentally examine the motion of a single particle in a Maxwell fluid as it transitions from ballistic to elastically trapped to diffusive motion, the first observation of this kind.  We compare these results to existing microscopic models for Maxwell fluids \cite{mason_optical_1995, van_zanten_brownian_2000, raikher_brownian_2013, grimm_brownian_2011}, and find significant discrepancies between the model predictions and the observed behavior.

An early effort to model the balistic diffusive transition was performed with the ideal gas approximation \cite{lemons_paul_1997, uhlenbeck_theory_1930}.  A more accurate model for dense fluids, motivated by early computer simulations \cite{rahman_liquid_1966,alder_decay_1970}, was achieved by adding an effective mass term and a memory term to the ideal gas model \cite{zwanzig_hydrodynamic_1970, hinch_application_1975}.  The effective mass term models the frictionally bound fluid that is attached to the particle and the memory term models the inertial interaction of the particle with nearby moving fluid \cite{zwanzig_compressibility_1975}.  At sufficiently short timescales and close to the speed of sound in the fluid this model breaks down and is replaced with the simple ideal gas model. The memory term in the dense fluid model, comes from the entrained fluid in a dense sytem which slows the change of direction.

These modifications to the Langevin equation were analytically solved \cite{clercx_brownian_1992} under the assumptions that the fluid is viscous and incompressible, the Reynolds number is low, and the test particle is a hard sphere \cite{widom_velocity_1971, l._d._landau_&_e.m._lifshitz_fluid_1959}.  The predicted MSD is included in the supplementary information and plotted in figure \ref{trla7Var}.   The dense fluid model and the ideal gas model both share similar asymptotic forms. At short times, known as the ballistic regime, the MSD asymptotes to $(2k_{B}T/M)t^2$. At long times, the diffusive regime, the MSD scales as $(4k_{B}T/\gamma)t$. Here $\gamma$ is the Stokes value ($6 \pi r \eta$), $k_B$ the Boltzman constant, $T$ the temperature, $\eta$ the viscosity, $r$ the tracer radius, $M$ the effective mass which is  $m+ \frac{1}{2}m_f$, where $m_f = \frac{4}{3} \pi r^3 \rho$ is the the mass of the fluid displaced by the colloid, and $\rho$ is the density of the fluid.  The dense fluid MSD differs from the ideal gas MSD in two salient ways: 1) It gives rise to a slower ballistic velocity, caused by the increased effective mass of the particle. 2) It has a much gentler crossover between ballistic and diffusive motion, caused by the inertial memory of the liquid.

Non-Newtonian fluids, however, have much more complicated Langevin equations \cite{van_zanten_brownian_2000, raikher_brownian_2013} which have not been explored as intensely due to the lack of experimental data at the shortest length and time scales.  One of the simplest non-Newtonian fluids is a Maxwell fluid, characterized by a single terminal relaxation time between spring like and viscous like behavior.  This requires the addition of a decaying spring term to the Langevin equation.  This additional term results in an intermediate plateau regime in the MSD corresponding to the behavior of a thermal spring.  While analytical solutions for the Maxwell fluid are lacking, predictions have been made about the asymptotic behavior in the three regimes: ballistic motion, elastically trapped motion, and finally diffusive motion.  The short-time ballistic behavior is predicted to asymptote to $(4k_{B}T/m)t^2$, the elastic trap should have a constant MSD of $2 k_{B}T/\pi r G_0$, and the long-time diffusive motion is predicted to have an asymptote of $(4k_{B}T/\gamma)t$ \cite{van_zanten_brownian_2000}. Here $G_0$ is the plateau modulus, a commonly measured rheological value which measures the amplitude of the storage and loss modulii \cite{aramaki_composition-insensitive_2010}. We use our technique to test the rheological predictions when applied to a single unconstrained particle in such a fluid moving between ballistic and diffusive regimes.

\begin{figure}[t]
\centerline{\includegraphics[scale=0.6]{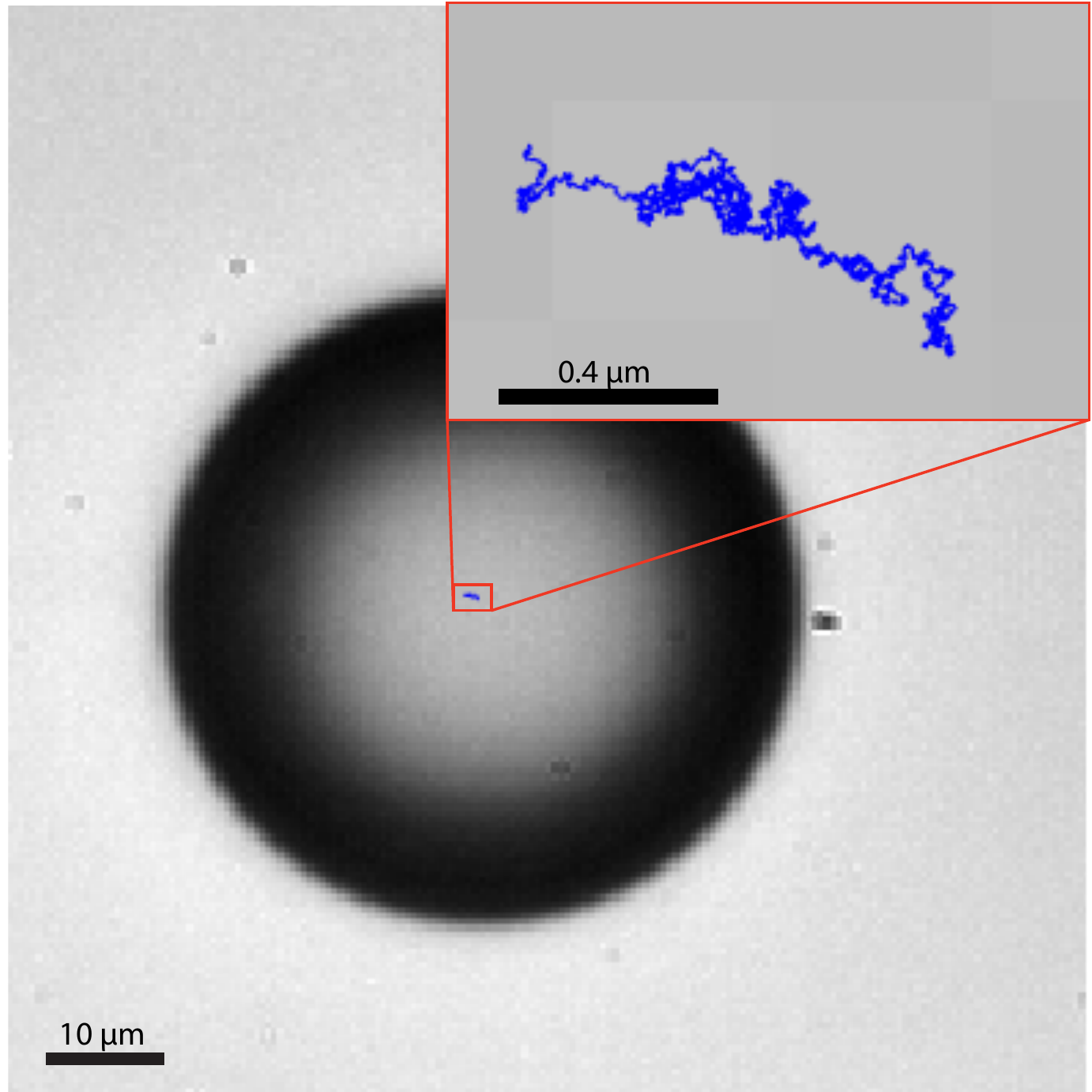}}
\caption{A single colloid suspended in water viewed through the microscope. The path the particle travels for the next 2 seconds is shown in blue.}
\label{particleTrace}
\end{figure} 


In this experiment, we used polystyrene hard spheres with a radius of 21.8 $\mu$m as our tracer particle.  Polystyrene was chosen because it is easily density matched to water using NaCl with only minimal, and known, changes to the viscosity. The size of the particle, larger than those in optical trap experiments, was chosen because the tracking precision as well as the ideal gas transition time and length increase with increasing radius.   We chose to use water as our experimental liquid because of its ubiquity in experiments and its relatively low viscosity. For our setup, the Reynolds number is $2.4\times10^{-9}$.  This system fulfills all of the underlying assumptions required by the  dense fluid equation.  When selecting a Maxwell fluid we chose to use a solution of Cetyltrimethylammonium chloride (CTAC) mixed with water.  This mixture has been found to exhibit a Maxwell fluid behavior caused by worm-like micelles \cite{oda_effect_1998, wilhelm_rotational_2003} and is a commonly studied Maxwell fluid. The micelles formed produce a network within the fluid.  This network, acting together, changes the properties of the supporting fluid.  Because the test particle we use is larger than the miccelles it will probe the properties of the complex fluid rather than solely those of the intervening fluid.

Test particles were placed into a deionized water mixture at $5\times10^{-3}$ \% w/v of colloid.  The colloids are slightly denser than water, so NaCl was added to density match the system at a measured value of $1.06 \times 10^{3}$ kg/m$^3$.  After sonicating and degassing the colloid-water solution, it was placed in a Fastwell silicon spacer cavity between a slide and a cover slip.  The chamber was sealed with vacuum grease to ensure that air bubbles did not form.  The silicon spacer had a width of 2.4 mm and circular void with a radius of 5 mm allowing the colloids to be imaged far from wall effects.  The slides were cleaned with piranha solution and dried with nitrogen gas, which removed any coatings on the slides.

We created a Maxwell fluid using a solution of CTAC at 1\% by weight with water with the addition of 0.12 M of NaSal to facilitate the formation of micelles.  We directly measured the plateau modulus of this solution to be 5.72 Pa using an angular frequency sweep from 62 - 0.062 rad/s at a constant displacement of 0.18 mrad (TA Instruments AR-2000ex rheometer, with a 60mm 1.025$^{\circ}$ cone plate).  Our fluid had a density of $1.055 \times 10^{3}$ kg/m$^3$, slightly lower than the average density of the beads.  However, the density is close enough that beads did not fall out of suspension until well after all the measurements were complete.  The same test particles were added at a concentration of  $2.5\times10^{-3}$ \% w/v.  When preparing samples, we used a process almost identical to the one for water.  The major difference was that the sample was not sonicated ahead of being added to the Fastwell because the fluid solidified when exposed to high frequency agitation.  Instead, the sample was slowly mixed using a low frequency mixer.

All data was collected on a Nikon TE2000s microscope on a floating stage optical table in a climate controlled room. Illumination was provided by a 500mW red LED (Thorlabs  LED635L) shining through the microscope condenser.  In between the LED and sample a neutral density filter on a swivel mount was added to allow initial setup to be done without excessive local heating of the sample.  The sample was encased in a small cardboard box for isolation from acoustic vibrations. Images were gathered through a 50x lens (Nikon LU plan ELWD 50x/0.55 B $\inf$/0 WD 10.1) using a Phantom M310 high speed camera.  Videos were taken at 40,000 fps (T = 25 $\mu$s) with an image size of 192x192 pixels and a magnification of 0.4$\mu$m/pix.  When filming a particle all motorized elements on the microscope and camera were turned off to eliminate small vibrations.   Once a particle was found, filming lasted 2.84 s (113,600 frames) after which the LED was immediately shut off and the ND filter replaced.

We used a radial center tracking algorithm\cite{parthasarathy_rapid_2012} to find the center of the colloids in progressive frames of the video. Using a combination of simulations and tracking test particles which were stuck to the slide, we found that the algorithm did not exhibit a preferred direction.  We found that the mean position error was about $1.5$ nm in each frame.  A representative trace with the first video frame is shown in figure \ref{particleTrace}.


We calculate the mean squared displacement (MSD) from our measured data as $\text{MSD}(\tau) = \langle \| \vec x(t+\tau)-\vec x(t) \| ^2 \rangle $ where $\vec x(t)$ is the measured position of the particle at time $t$, $\tau$ is the lag-time between position measurements, and angle brackets denote a time average.  The MSD for a representative particle is plotted as green squares in figure \ref{trla7Var}.  This MSD exhibits a small drift at long times, past about 0.1 s, and a noise floor at very short times.  The drift is likely the result of convective flows within our sample chamber, driven, perhaps, by local heating of the sample.  However, this drift can be easily removed by calculating the variance of particle position as a function of lag time as $\text{Var}(\tau) = \text{MSD}(\tau)-\| \langle \vec x(t+\tau)- \vec x(t) \rangle ^2 \|$.  The noise floor is caused by photon shot noise in our camera contributing to uncertainty in the localization of a particle.  We can directly measure this noise by tracking a particle fixed to a slide and find it to be independent and identically distributed Gaussian noise with a variance of approximately $2\times10^{-18}$ m$^2$ (inset to figure \ref{trla7Var}).  The precise value of this noise variance changes from run to run due to variations in particle size and particle focus (due to changes in z-position).  Because this noise is independent and identically distributed for a given measurement we can simply subtract the noise floor from our measurement to find the true variance of our particle, plotted as black circles in figure \ref{trla7Var}.

\begin{figure}[t]
\centerline{\includegraphics[scale=1.1]{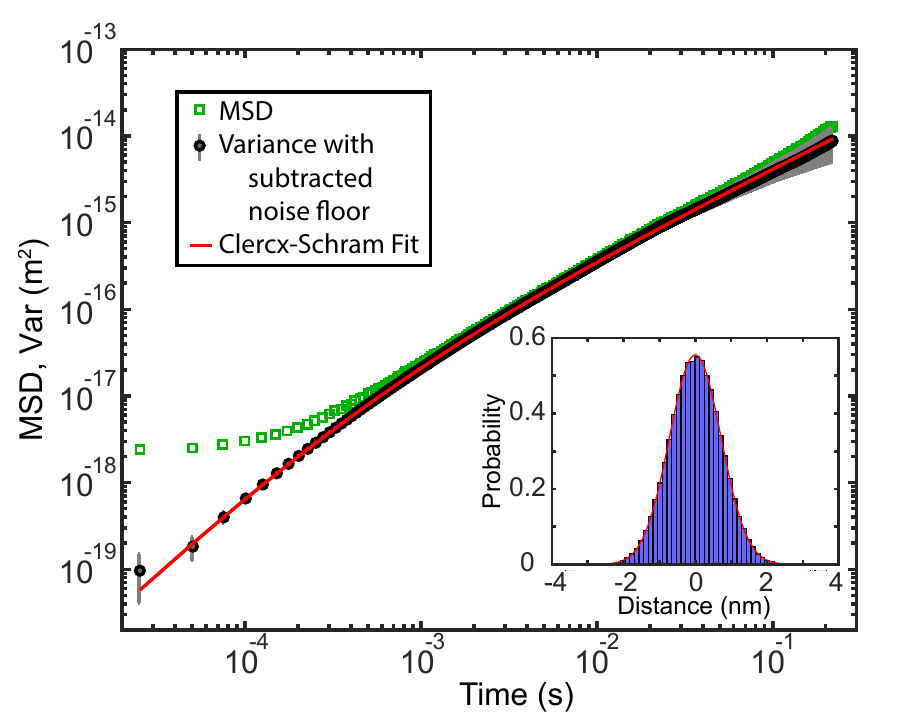}}
\caption{ The variance of the colloid's trace, shown in figure~\ref{particleTrace}.  The green squares show the mean squared displacement. The black dots with the error bars show the variance with the noise floor subtracted, in this case $2.34\times10^{-18}$ m$^2$.  In red is the fitted Clercx-Schram theoretical prediction for the MSD.  The best fit value for Temperature is 302 K, and for radius $21$ $\mu$m.  We propagate the localization error of the position measurement as well as sampling error through our calculation to obtain error bars for the plot as described in the supplementary material.  Inset: A histogram of measured positions for a stranded particle.}
\label{trla7Var}
\end{figure} 

The plotted variance clearly shows a ballistic regime below about $10^{-3}$ to $10^{-4}$ s, a crossover regime up till about $10^{-2}$ s, and a diffusive regime for longer times.  The measured variance fits the dense fluid model exceedingly well over the entire range of measured lag-times as shown in figure \ref{trla7Var}.  The model depends on four physical parameters: 1) temperature, 2) particle radius, 3) fluid density, and 4) fluid viscosity.  Of these, we independently measure the fluid density prior to observation.  The fluid viscosity of salt water is a known function of density and the temperature \cite{sharqawy_thermophysical_2010}.  Therefore, we have only two independent fitting parameters: temperature and particle radius.  To this, we add a third fitting parameter to describe the magnitude of the noise floor.

\begin{figure}[t]
\centerline{\includegraphics[scale=0.70]{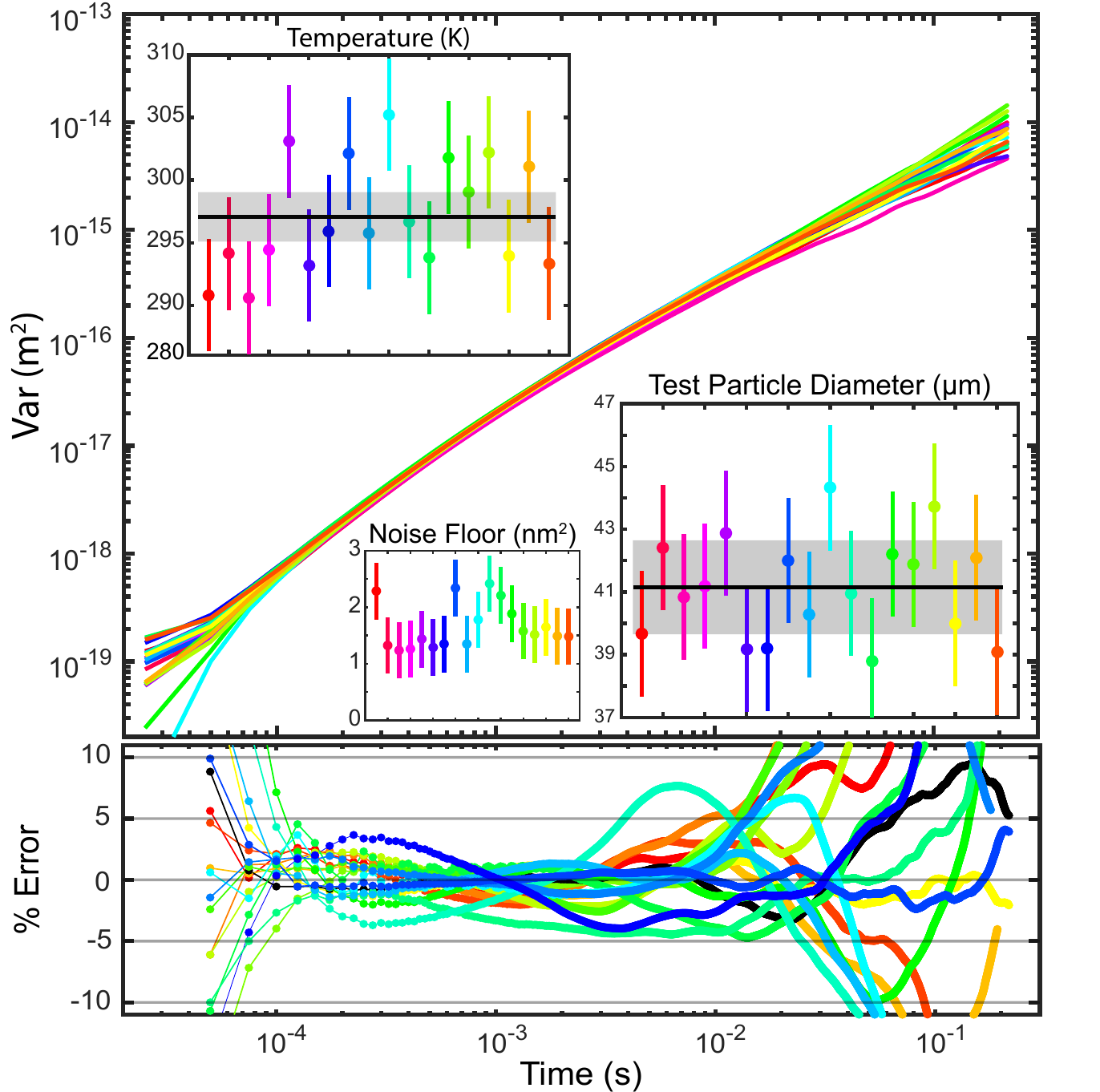}}
\caption{Top: The noise floor adjusted variance for 18 different videos. Inset: The fitted temperature, test particle radius, and noise floor for the different trials.  The black line shows the mean values with a gray standard deviation.  Bottom: Residual percentages showing by what percent the measurement deviates from the individual fits for each of the corresponding variance's shown above.}
\label{residpPlot}
\end{figure}

We independently fit 18 measurements using 18 different particles, shown in figure \ref{residpPlot}.  On average, the particle radius was found to be $20.5 \pm 0.8$ $\mu$m, within tolerance of the manufacturer's quoted radius.  The average temperature measured by our fitting was found to be slightly higher ($297 \pm 4.5$ $K$) than the measured room temperature ($293 \pm 2$ K), likely the result of local heating from the intense illumination.  The noise floors for the measurement were found to range from $1.2\times10^{-18}$ m$^2$ to $2.4\times10^{-18}$ m$^2$.  Thus fit, the dense fluid functional form is indistinguishable from the data over much of our measured range.  To characterize the agreement, we plot the residual percentages, and find them to be unbiased and with error less than 5\% over at least two decades of lag time, as shown in figure \ref{residpPlot}.  At longer times, where drift and sampling errors increase, the percent error increases as well.

\begin{figure}[t]
\centerline{\includegraphics[scale=1.05]{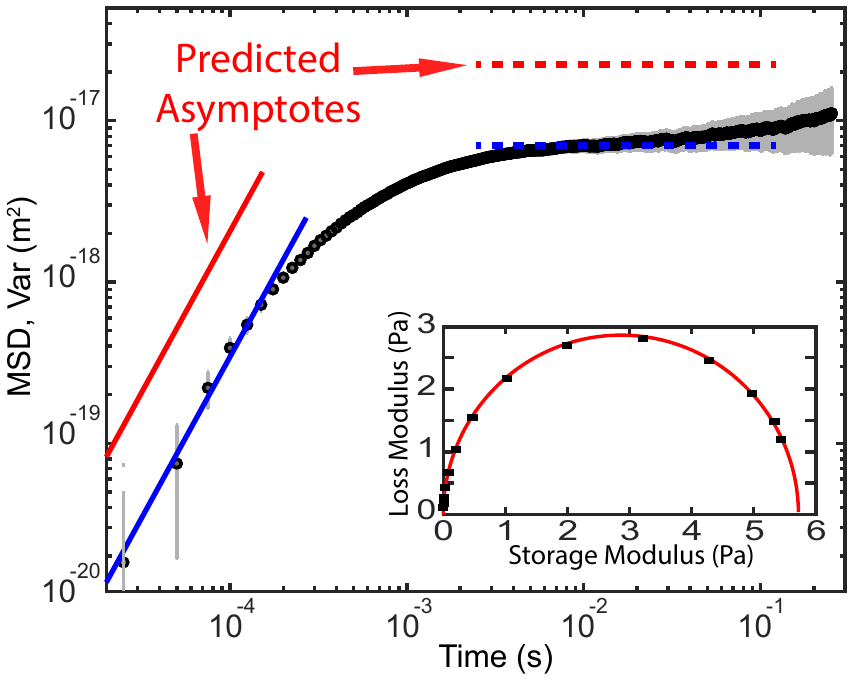}}
\caption{The MSD and adjusted variance for a particle moving in a Maxwell fluid.  The black dots show the noise floor subtracted variance.  In this figure the noise floor used is $2.2883\times10^{-18}$ m$^2$. The red lines show the predicted theoretical asymptotes \cite{van_zanten_brownian_2000} for the ballistic (solid) and plateau (dashed) regimes.  The blue lines show the observed asymptotes for the ballistic (solid) and plateau (dashed) regimes.  Inset: The Cole-Cole plot for CTAC as measured with a conventional rheometer.  The red curve is a fit to the Maxwell function demonstrating a measured plateau modulus of 5.72 Pa.}
\label{ctacTrl5msd}
\end{figure}

We perform similar experiments in a Maxwell fluid created with a solution of CTAC and water as described above.  As in the case with water, we see a minimum noise floor at short times and a long time drift in the MSD.  The drift in the measurement is removed by using the variance as described above and the noise floor is estimated and subtracted as shown in Figure \ref{ctacTrl5msd} for a representative trial.  In total 30 independent measurements were made with this Maxwell fluid.

The plotted variance for a Maxwell fluid has two notable features.  1) At short times the motion is clearly ballistic.  The best fit prefactor for the asymptote is however considerably lower than the one predicted by either the ideal gas or dense fluid models, corresponding to an effective mass six times larger than the particles mass or an entrained region with a radius 39.6$\mu m$, compared to the reported radius of 21.8$\mu m$.  This increased effective size of the particle can perhaps be understood as a result of the fact that the surrounding fluid contains a network of worm-like micelles.  The test particle impinges upon the network of intertwined micelles and pulls some of them along thus increasing the particle's effective mass.   Alternatively, the surface of the particle may actually attract the micelles which would increase the effective mass as well.  However, due to the presence of salt in this solution any interaction between the particle and the micelles must neccessarily be small.  2) The variance shows a clear secondary plateau which is independent of the noise floor.  This plateau is characteristic of thermally damped motion consistent with a Maxwell fluid's predicted behavior for high frequencies.  Examining the best fit asymptotes to the plateau regime, we find an average plateau modulus over all measurements of 16 $\pm$ 2 $Pa$.  This is almost a factor of 3 larger than the rheometer measured value of 5.7 $Pa$.  

The Cole-Cole plot (inset to Figure \ref{ctacTrl5msd}) for this fluid shows it to be a perfect Maxwell fluid when measured on a conventional rheometer, however this deviates from the observed microscopic behavior.  These results demonstrate that at the short time and length scales that our technique probes, the physics governing this fluid are in fact significantly more complicated than those of a simple Maxwell fluid model.  The displacement scales probed with our technique are just under 4 orders of magnitude smaller than those accessible to a rheometer and the time scales are two orders of magnitude smaller.


In this experiment, we resolve the functional form of the ballistic crossover, revealing the fundamental length and time scales between individual and collective interactions in both Newtonian and Maxwell fluids.  In so doing we have created a microscale first-principles thermometer based on the kinetic theory definition of temperature.  We have demonstrated the validity of this approach by the extremely precise agreement between our results and theoretical models for motion in dense Newtonian fluids.  We have experimentally tested the accuracy of Maxwell fluid Langevin equation solutions and found them to be wanting in accurately describing real materials.  Asymptotically, we see a clear need for the addition of an effective mass term.  More troublingly, the plateau values as measured with this method are markedly different from those found with a conventional rheometer.  This difference may be a sign of a shift in behavior between the microscale addressed by our measurement and the macroscale measurement performed with a rheometer, suggesting that materials which appear Maxwell at large length scales may be more complicated at small lengthscales.  Alternatively, this result could be indicative that the assumptions used in deriving the asymptotic behavior of the model need to be further modified.  Our technique provides an independent method for testing models for the microscopic structure of fluids and the accompanying macroscopic fluid constants.  In the future, this method promises to be useful in measuring multiple transitions between motion regimes in viscoelastic materials, an area where laser traps have difficulty because of the effects of confinement \cite{grimm_brownian_2011}.  This method will also enable detailed studies of the influence of long range interactions, such as wall effects, in an interaction free manner \cite{felderhof_effect_2005}.  As such, high speed single particle tracking promises to become an important tool in the study of the fundamental behavior of liquids.



We thank Raghu Parthasarathy, Tristan Ursell, and Mike Taormina for helpful discussions, Travis Walker and Britany Swann for help with the rheometer, and the University of Oregon machine and electrical shop staff.  This work was supported by National Science Foundation (NSF) Career Award DMR-1255370.


\bibliography{BallisticDiffusive2Bib.bib}

\end{document}